\documentstyle[11pt,epsf]{article}
\newcommand{\beq}{\begin{equation}}
\newcommand{\eeq}{\end{equation}}
\newcommand{\bea}{\begin{eqnarray}}
\newcommand{\eea}{\end{eqnarray}}
\newcommand{\bi}{\bibitem}

\begin{document}

\title{ QCD Thermodynamics with an Improved Lattice Action }

\author{Claude Bernard, James E.\ Hetrick
\\
{\it Department of Physics, Washington University, St.~Louis, MO 63130, USA}
\and
{Thomas DeGrand, Matthew Wingate}
\\
{\it Department of Physics, University of Colorado, Boulder, CO 80309, USA}
\and
{Carleton DeTar}
\\
{\it Department of Physics, University of Utah, Salt Lake City, UT 84112, USA}
\and
{Steven Gottlieb}
\\
{\it Department of Physics, Indiana University, Bloomington, IN 47405, USA}
\and
{Urs M.\ Heller}
\\
{\it SCRI, Florida State University, Tallahassee, FL 32306-4052, USA}
\and
{Kari Rummukainen}
\\
{\it Fakult\"at f\"ur Physik, Universit\"at Bielefeld,
D-33615, Bielefeld, Germany}
\and
{Doug Toussaint}
\\
{\it Department of Physics, University of Arizona, Tucson, AZ 85721, USA}
\and
{Robert L.\ Sugar}
\\
{\it Department of Physics, University of California, Santa Barbara, CA 93106, USA}
\and
{(MILC Collaboration)}
}

\date{\today}
\maketitle

\begin{abstract}
We have investigated QCD with two flavors of degenerate fermions
using a Symanzik-improved lattice action for both the gauge and fermion
actions.  Our study focuses on the deconfinement transition
on an $N_t=4$ lattice.  Having located the thermal transition,
we performed zero temperature simulations nearby in order to 
compute hadronic masses and the static quark potential.  We find
that the present action reduces lattice artifacts present in
thermodynamics with the standard Wilson (gauge and fermion) actions.
However, it does not bring studies with Wilson-type quarks to the
same level as those using the Kogut--Susskind formulation.

COLO-HEP-381,
FSU-SCRI-97-17
\end{abstract}

\newpage

\section{Introduction}
\label{sec:intro}

An understanding of the high temperature behavior of QCD is desirable in 
addressing problems such as heavy ion collisions and the evolution of
the early universe.  It is believed that, at a
temperature between $140-200$ MeV (where pions are produced copiously),
hadronic matter undergoes
a transition to a plasma of quarks and gluons.  This phenomenon is
intrinsically nonperturbative, and Monte Carlo lattice simulation
provides the best theoretical tool with which to study it.

Most lattice studies have used Kogut--Susskind (KS) fermions because of
their exact U(1) chiral symmetry at finite lattice spacing.  The full
SU(2) chiral symmetry is recovered in the continuum limit. Although in
contrast Wilson fermions explicitly break chiral symmetry,
the continuum limit of the different discretizations
is expected to be the same.  One advantage of simulating with two
flavors of Wilson quarks versus two flavors of KS fermions is that
the updating algorithm is exact in the former case but has finite
time step errors in the latter.  The price to be paid is that
the explicit chiral symmetry breaking gives rise to an additive mass 
renormalization; thus, the location of the chiral limit is not 
known {\it a priori}.

Even more troublesome for dynamical
Wilson fermions is the presence of lattice artifacts which
qualitatively affect physics at large lattice spacing.  In
Refs.~\cite{ref:MILC4} and~\cite{ref:MILC6} it was found that the 
deconfinement transition becomes very steep for
intermediate values of the hopping parameter, $\kappa$. 
In fact, on an $N_t=6$
lattice the transition appears to be first order for a range of intermediate
hopping parameters and smooth otherwise.

The lattice community has worked very hard recently to construct
actions which have fewer lattice artifacts than the standard
discretizations of the continuum action.  One philosophy is
to add operators to the action which cancel ${\cal O}(a^n)$ terms
in the Taylor expansion of spectral observables.
It is plausible that an action which converges to the
continuum action faster in the $a \rightarrow 0$ limit would be
free of the artificial first order behavior.  We adopt this improvement
program, attributed to Symanzik, in the present work.

An alternative approach is to search for an action which lies
on or near the renormalized trajectory of some renormalization
group transformation.  Since all irrelevant couplings are zero along the
renormalized trajectory, actions there have no scaling violations,
i.e.\ are quantum perfect.  Such an action was approximated by
Iwasaki~\cite{ref:IWASAKI_RG} and has being used to study QCD
thermodynamics with two flavors of unimproved Wilson
fermions~\cite{ref:TSUKUBA}.  Although it is still an open
question to what extent this action lies on a renormalized
trajectory, the results of Ref.~\cite{ref:TSUKUBA} show
improvement over standard Wilson thermodynamics.

In this paper we report on our simulation of finite temperature
lattice QCD with two flavors of ${\cal O}(a)$ Symanzik-improved
fermions and ${\cal O}(a^2)$ Symanzik-improved glue.  We describe
the action we used in Section~\ref{sec:action}.
In Section~\ref{sec:simul} we give the details of our
simulations, and we present our results in Section~\ref{sec:results}.
Finally, we give our conclusions in Section~\ref{sec:concl}.

\section{Action}
\label{sec:action}

\begin{figure}
\vspace{1.6in}
\includegraphics{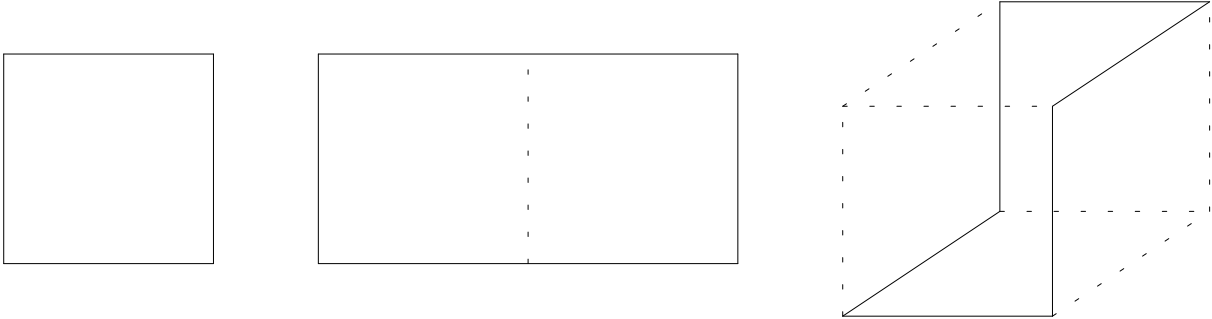}
\caption{The three Wilson loops in the one-loop Symanzik
improved gauge action that we used. }
\label{fig:gauge}
\end{figure}

When one expands a lattice operator in a Taylor series about zero
lattice spacing $a$, one recovers its relevant (or marginal)
continuum operator plus higher dimensional irrelevant operators
proportional to powers of $a$.  Symanzik suggested that by
selecting a favorable combination of lattice operators in the
lattice action, one might have cancellations of the irrelevant
operators up to some order in the lattice spacing~\cite{ref:SYMANZIK}.
L\"uscher and Weisz have applied this philosophy to SU($N$)
gauge theories.  They imposed an on-shell improvement condition
whereby discretization errors are eliminated order-by-order
in $a$ from physical observables and constructed an ${\cal O}(a^2)$
improved gauge action~\cite{ref:LW}.
Furthermore,
they computed the coefficients of the operators in the action
through one-loop order in lattice perturbation theory~\cite{ref:LWPURE}.
This improvement condition does not provide a unique
action.  The choice which is the most efficient in terms of
computational effort adds a $1 \times 2$ rectangle and a 6-link 
``twisted'' loop to the Wilson plaquette action.
(See Fig.~\ref{fig:gauge}.)

It is well-known now that lattice perturbation theory in the
bare coupling $g_0$ is not trustworthy.  The $a^2$ in the vertex
of the tadpole graph is cancelled by the ultraviolet divergence of
the gluon loop.  Therefore, hidden in the higher order terms
of the expansion in $a$ are tadpole graphs which give an effective
$a^0\sum_n c_n g^{2n}$ contribution.
A standard way to deal with this problem is to
define a mean link $u_0$ and replace $U_\mu \rightarrow
U_\mu/u_0$~\cite{ref:PARISI,ref:LM}.  This introduces a ``boosted'' coupling
constant $g^2 = g_0^2/u_0^4$.

Here we combine these two ideas, Symanzik improvement of the action
and ``tadpole improvement'' of lattice perturbation theory.
Our gauge action for this work is as derived in Ref.~\cite{ref:ALFORD},
\bea
S_g & = & \beta \sum_{\rm plaq} {1\over3}~{\rm Re~Tr~}(1 - U_{\rm plaq})
\nonumber \\
& + & \beta_1 \sum_{\rm rect} {1\over3}~{\rm Re~Tr~} (1 - U_{\rm rect}) \nonumber \\
& + & \beta_2 \sum_{\rm twist} {1\over3}~{\rm Re~Tr~} (1 - U_{\rm twist}),
\label{eq:gaction}
\eea
where $\beta$ is a free parameter 
(in this normalization $\beta = {6\over g^2 u_0^4}{5\over3}
(1 - 0.1020 g^2 + {\cal O}(g^4))$), and
\bea
\beta_1 = -{\beta \over 20 u_0^2}~
\bigg[1 + 0.4805~\bigg({g^2\over4\pi}\bigg)\bigg]
& = & -{\beta \over 20 u_0^2}~\Big(1 - 0.6264~\ln(u_0)\Big) 
\label{eq:beta1} \\
\beta_2 = -{\beta \over u_0^2}~0.03325~\bigg({g^2\over4\pi}\bigg)
& = & {\beta \over u_0^2}~0.04335~\ln(u_0).
\label{eq:beta2}
\eea
The subscripts ``plaq'', ``rect'', and ``twist'' refer to
the $1 \times 1$ plaquette, the planar $1 \times 2$ rectangle,
and the ``$x, y, z, -x, -y, -z$'' loop, respectively.
Following Ref.~\cite{ref:LM} we have chosen to define
the mean link $u_0$ through
\beq
u_0 ~\equiv~ \Big({1\over3}~{\rm Re~Tr~}\langle U_{\rm plaq} 
\rangle\Big)^{1/4},\label{eq:link}
\eeq
and the strong coupling constant is defined through the 
perturbative expansion of the plaquette~\cite{ref:WW}
\beq
{g^2\over4\pi} ~\equiv~ -{\ln\Big({1\over3}~{\rm Re~Tr~}\langle U_{\rm plaq} 
\rangle\Big) \over 3.06839}.
\label{eq:alphas}
\eeq
In the Monte Carlo simulations, we tune $u_0$ in the action to
be consistent with the fourth root of the average plaquette.
This procedure is discussed in more depth in Section~\ref{sec:simul}.

\begin{figure}[hb]
\vspace{2.6in}
\includegraphics{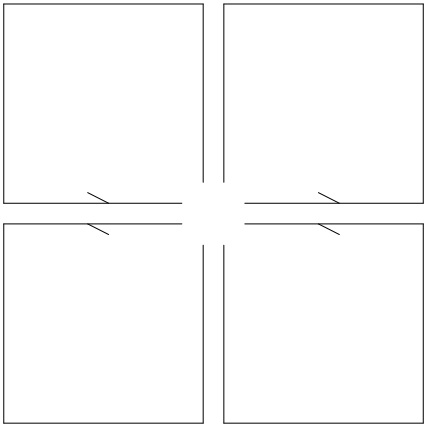}
\caption{The clover term in the Sheikholeslami--Wohlert action. }
\label{fig:clover}
\end{figure}

The Wilson fermion action has errors of ${\cal O}(a)$.  The
Symanzik improvement program can be extended to improving
this action too.  In the $a \rightarrow 0$
limit, the ${\cal O}(a)$ term in the action is proportional
to $\overline{\psi}D^2\psi$ and can be removed by adding
a next-nearest neighbor operator to the action \cite{ref:HW},
or, after an isospectral transformation of the fermion fields,
by adding a magnetic interaction \cite{ref:SW}.  Thus, the 
tree-level Symanzik improved action is
\beq
S_f ~=~ S_W - {\kappa \over u_0^3} \sum_x \sum_{\mu < \nu} \bigg[ 
\overline{\psi}(x) ~ i \sigma_{\mu\nu} F_{\mu\nu} \psi(x) \bigg],
\eeq
where $S_W$ is the usual Wilson fermion action, and
\beq
i F_{\mu\nu} = {1\over8} ( f_{\mu\nu} - f_{\mu\nu}^\dagger ).
\eeq
$f_{\mu\nu}$ is the clover-shaped combination of links
(see Figure~\ref{fig:clover}).
As with the gauge action, the links here are also
tadpole improved.  Note that one factor of $u_0$ is absorbed into the
hopping parameter.

Both these gauge and fermion actions have been used widely,
e.g.\ in studies of finite temperature
SU(3)~\cite{ref:IMPSU3THERMO,ref:IMPSU3THERMO2}
and quenched spectroscopy~\cite{ref:UKQCD_CLOV,ref:SCRI_LAT96,ref:SCRI_CLOV}.
At least one group is in the progress of using this action
to calculate spectroscopy on unquenched configurations~\cite{ref:EDWARDS}.
Therefore,
we believe our choice of action to be well-justified and useful
for comparison to other work.

Of course, further progress is being made in refining the Symanzik
improvement program.  One can attempt to set the coefficients of the
higher dimension operators nonperturbatively by demanding that,
for example, Ward identities be satisfied up to some order in
the lattice spacing~\cite{ref:SYM_PCAC}.  Also, fermion 
actions which are constructed to have errors of ${\cal O}(a^3)$ to
${\cal O}(a^4)$ are currently being tested~\cite{ref:D234,ref:WOLOSHYN}.

\section{Simulation details}
\label{sec:simul}

Our finite temperature simulations were on an $8^3\times4$
lattice.  At fixed $\beta$ (= 6.4, 6.6, 6.8, 7.0, 7.2, 7.3, and
7.4) we varied $\kappa$ in small increments across the crossover.
Our microcanonical time step was such
that the acceptance rate was between 60-80\%; typically 
$\Delta t = 0.03$, but for the stronger gauge coupling (lighter quark
mass) runs $\Delta t = 0.01$.  We accumulated over 1000 trajectories
at points close to the crossover.
For the updating, we used the standard hybrid molecular
dynamics (HMD) algorithm~\cite{ref:HMD} followed by a
Monte Carlo accept/reject step
(hybrid Monte Carlo, or HMC)~\cite{ref:HMC}.  The calculation
of the ``clover'' contribution to the HMD equations of motion
is tedious but 
straightforward.\footnote{Independent of our work, calculations of the
HMD equations of motion for this improved Wilson fermion
action have been published in~\cite{ref:HMD_SW}.}
The even-odd preconditioning technique developed for 
standard Wilson fermions~\cite{ref:ILU} is also implemented for
the improved Wilson fermion action~\cite{ref:ILU_SW}.

Our companion zero temperature runs were performed
on an $8^3\times16$ lattice at five
$(\beta,\kappa)$ points along the crossover, as well as at five
other points neighboring the crossover line.  Some of these
runs were extended in order to be able to extract the heavy
quark potential from Wilson loop expectation values.  As with
the finite temperature runs, we tried to maintain a Monte
Carlo acceptance rate around 60-80\%, so our time step varied
from $\Delta t = 0.005$ to 0.03.  Measurements of hadron
correlators and Wilson loops were taken every 10 trajectories.

The large majority of our simulations were performed on
the IBM SP2 at the Cornell Theory Center, and two finite
temperature simulations were run on a cluster of IBM RS/6000's at
the Supercomputer Computations Research Institute of
Florida State University.

We used the conjugate gradient (CG) matrix inversion algorithm to
compute $(M^\dagger M)^{-1}$ with a maximum residue of $10^{-6}$
during the HMC updating.  For the spectroscopy calculations,
where we wish to invert $M$, we found the stablized biconjugate
gradient (biCGstab) algorithm to be twice as efficient as
CG~\cite{ref:FROMMER_LAT96}.

We tuned $u_0$ so that it agreed with the fourth root of the space-like
plaquettes that we measured.  It might have been preferable to
do this on $T=0$ configurations, however, the heavy cost of repeatedly
equilibrating a $N_t=16$ lattice forced us to perform this tuning
procedure on the $N_t=4$ configurations.
In the next section we will show that the difference in the two ways
of tuning $u_0$ is small.
This tuning procedure would be dangerous if the system underwent a
first order phase transition, but, we will also show that
the plaquette varies smoothly across the transition.  Let us remark
that this tuning procedure is a prescription.  $u_0$ may be defined
in a number of ways since it is an {\it estimate} of the higher order 
tadpole contributions to perturbation theory calculations.  Therefore,
while one might argue that tuning $u_0$ strictly on a zero temperature
lattice would better estimate the tadpole contributions, our method is
well-defined and self-consistent.

\section{Results}
\label{sec:results}

\subsection{Thermodynamics}
\label{ssec:thermo}

The first task was to locate the thermal crossover line
$\kappa_T(\beta)$.  To this end we measured the expectation
values of the Polyakov loop, the plaquette, the quark
condensate $\langle\bar\psi\psi\rangle$,
and the number of CG matrix inversion
iterations as we generated the $8^3 \times 4$ configurations.

\begin{figure}[hb]
\vspace{3.3in}
\includegraphics{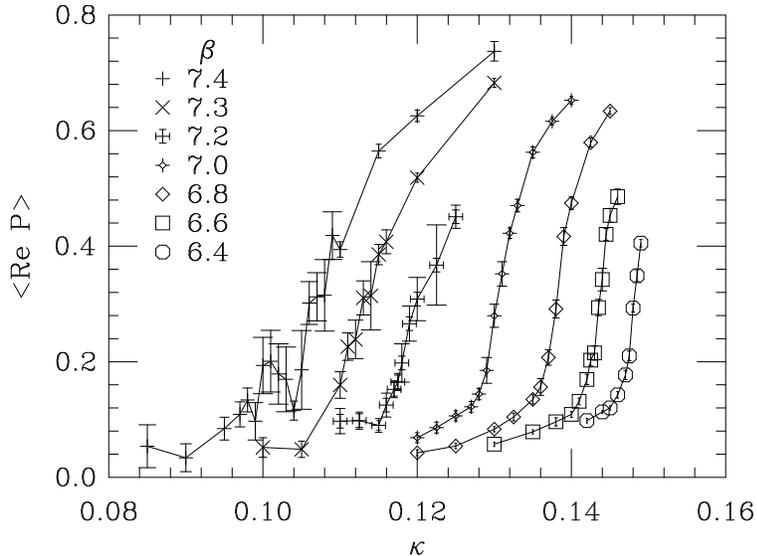}
\caption{Polyakov loop vs.~hopping parameter -- all $\beta$'s. }
\label{fig:pl_vs_k}
\end{figure}

\begin{figure}
\vspace{3.25in}
\includegraphics{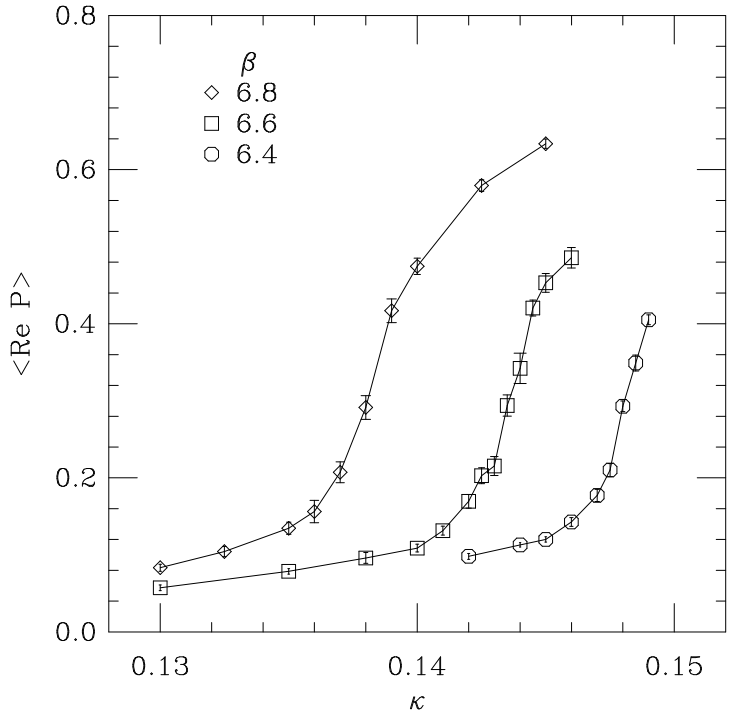}
\caption{Polyakov loop vs.~hopping parameter -- for the 3 lowest $\beta$'s. }
\label{fig:pl_vs_k2}
\end{figure}

In pure gauge theory the Polyakov loop is an order parameter
for the deconfinement transition: $\langle P\rangle = 0$ in
the confined phase because the free energy for a single color
triplet charge is infinite, while in the deconfined phase
the test charge can be screened, so the free energy is finite and
$\langle P\rangle \ne 0$.  For unquenched QCD, the Polyakov loop
is not an order parameter since it is nonzero even in the hadronic
phase, but it does increase dramatically at the transition.
In this work, we identify the thermal crossover as the place
where the derivative of $\langle{\rm Re}~P\rangle$ is greatest.
Figure~\ref{fig:pl_vs_k} shows $\langle{\rm Re}~P\rangle$
versus the hopping parameter $\kappa$ for the seven values of
fixed coupling $\beta$; Figure~\ref{fig:pl_vs_k2} shows only
the runs where the crossover is at the lowest 3 values of
$M_{\rm PS}/M_{\rm V}$.  Although the crossover becomes
steeper at stronger coupling, there is no evidence of a
first order transition.

\begin{figure}
\vspace{3.25in}
\includegraphics{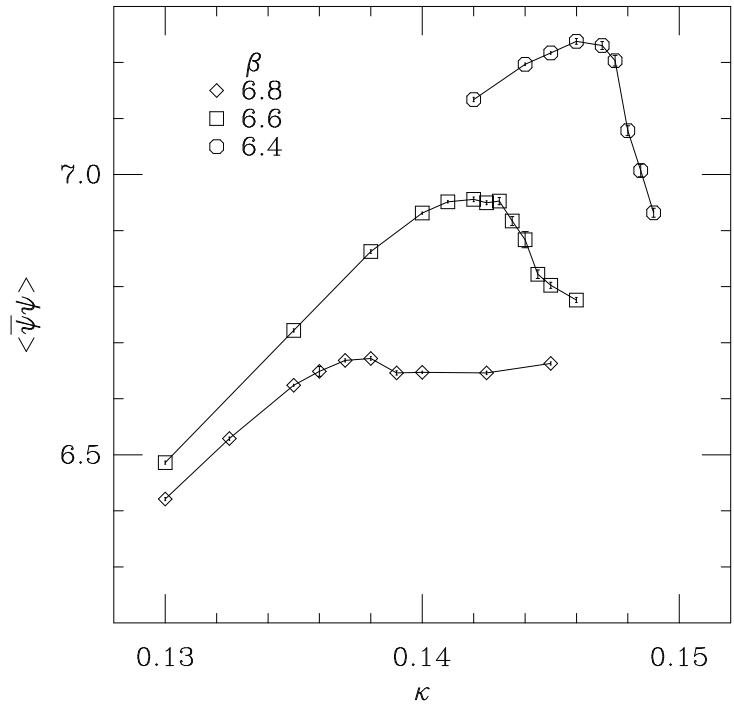}
\caption{$\langle\bar\psi\psi\rangle$ vs.\ hopping parameter. }
\label{fig:pbp_vs_k2}
\end{figure}

In continuum QCD with massless quarks, one expects to see
a restoration of the spontaneously broken chiral symmetry
at high temperatures.  The order parameter for this transition
is the chiral condensate $\langle\bar\psi\psi\rangle$.
Since Wilson fermions break chiral symmetry explicitly, the
meaning of $\langle\bar\psi\psi\rangle$ at $\kappa\ne\kappa_c$
is not so clear.  Besides the usual multiplicative renormalization
one must make a subtraction to compensate for the additive
renormalization of the quark mass.  A properly subtracted
$\langle\bar\Psi\Psi\rangle$ can be defined through an axial
vector Ward identity~\cite{ref:BOCHICCHIO}.
However, since our study did not
include calculation of screening propagators, we can only
look at the unrenormalized $\langle\bar\psi\psi\rangle$.
In spite of these problems, Figure~\ref{fig:pbp_vs_k2} shows
a drop in $\langle\bar\psi\psi\rangle$ at the crossover
identified by $\langle{\rm Re}~P\rangle$.

\begin{figure}
\vspace{3.25in}
\includegraphics{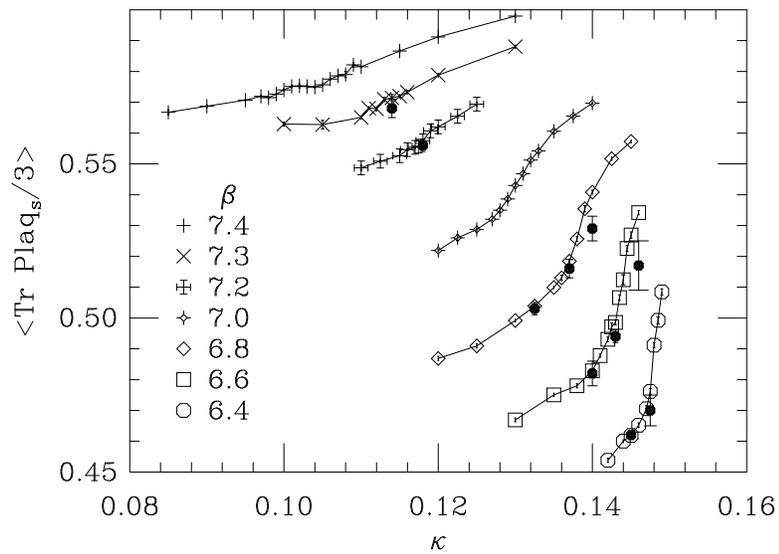}
\caption{Space-space plaquette vs.\ hopping parameter -- all $\beta$'s. 
The shaded octagons mark the zero temperature values.}
\label{fig:ssplaq_vs_k}
\end{figure}

\begin{figure}
\vspace{2.65in}
\includegraphics{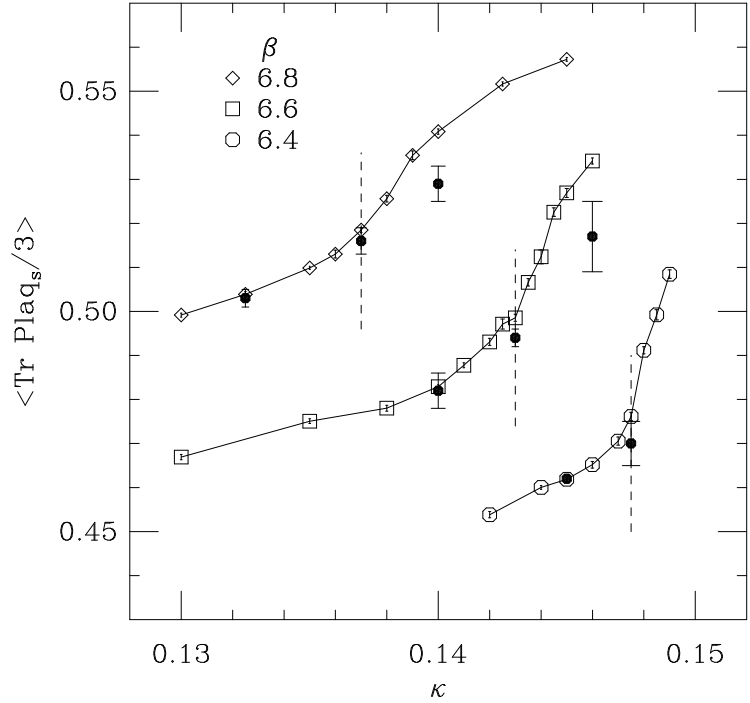}
\caption{Space-space plaquette vs.\ hopping parameter -- for the 3 lowest 
$\beta$'s.  The shaded octagons mark the zero temperature values.
The dashed lines indicate our determination of $\kappa_T(\beta)$
and emphasize the agreement between the space-like plaquettes measured
on zero temperature and finite temperature lattices at the crossover.}
\label{fig:ssplaq_vs_k2}
\end{figure}

\begin{figure}
\vspace{3.85in}
\includegraphics{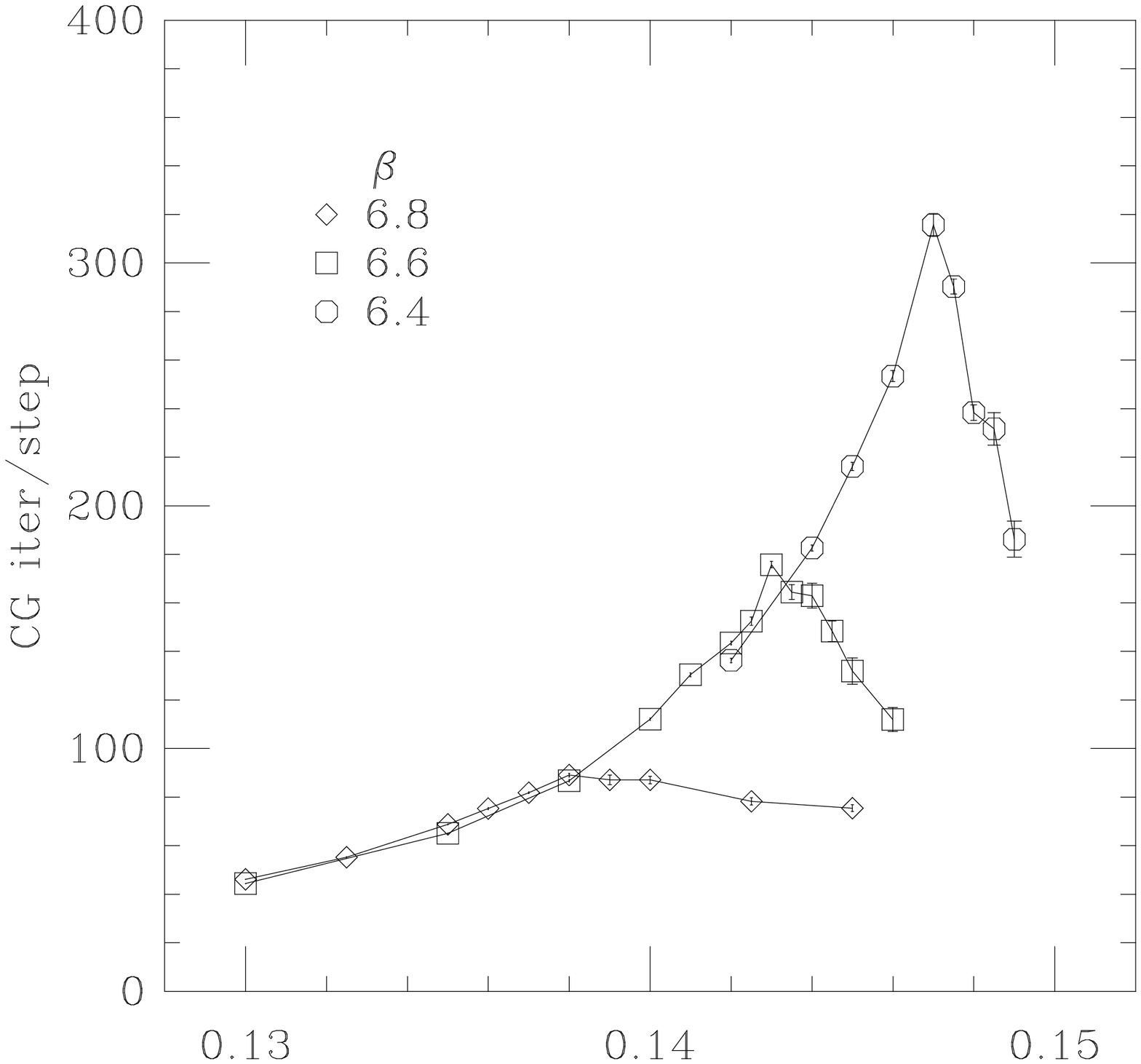}
\caption{Conjugate gradient matrix inversions vs.\ hopping parameter. }
\label{fig:cg_vs_k2}
\end{figure}

Since we use the plaquette (in the space-space planes) to
self-consistently tune $u_0$, we must ensure that it varies
smoothly across the thermal crossover.
Figures~\ref{fig:ssplaq_vs_k} and \ref{fig:ssplaq_vs_k2}
show that this is the case.  In fact,
the plaquette on the zero temperature lattices agrees within errors
with the plaquette at finite temperatures on the confined side of
the crossover.  The dashed vertical lines in those figures simply
mark the location of the crossover, $\kappa_T(\beta)$.
The large errors on the deconfined side are due to
the smaller sample sizes where running at lower quark mass is expensive.

As the thermal crossover line $\kappa_T(\beta)$ approaches
the critical line $\kappa_c(\beta)$, the number of iterations
needed to invert the fermion matrix per time step, 
$N_{\rm iter}$, peaks at
the thermal crossover.  The reason is that as one approaches
$\kappa_T(\beta)$ from the confined side (varying $\kappa$
with $\beta$ fixed) the zero modes at $\kappa_c(\beta)$
become more influential, while there are no zero modes in
the deconfined phase.  Figure~\ref{fig:cg_vs_k2} shows the
peaks in $N_{\rm iter}$ are at the same locations as the crossovers
indicated by the Polyakov loop.

\subsection{Spectrum}
\label{ssec:spectrum}

For a number of reasons, it is useful to evaluate some
zero temperature quantities at the parameters of our
thermodynamic simulations.  The light hadron spectrum
is essential in determining the chiral limit for Wilson-like
fermions.

The spectroscopy was an entirely straightforward lattice
computation which used Gaussian-smeared source wavefunctions and
point-like sink wavefunctions.  We performed correlated fits to a
single exponential
and selected the best fits based on a combination of smallest chi-squared per
degree of freedom and largest confidence level.  Propagators
are separated by 10 HMC trajectories.

\begin{figure}
\vspace{2.25in}
\includegraphics{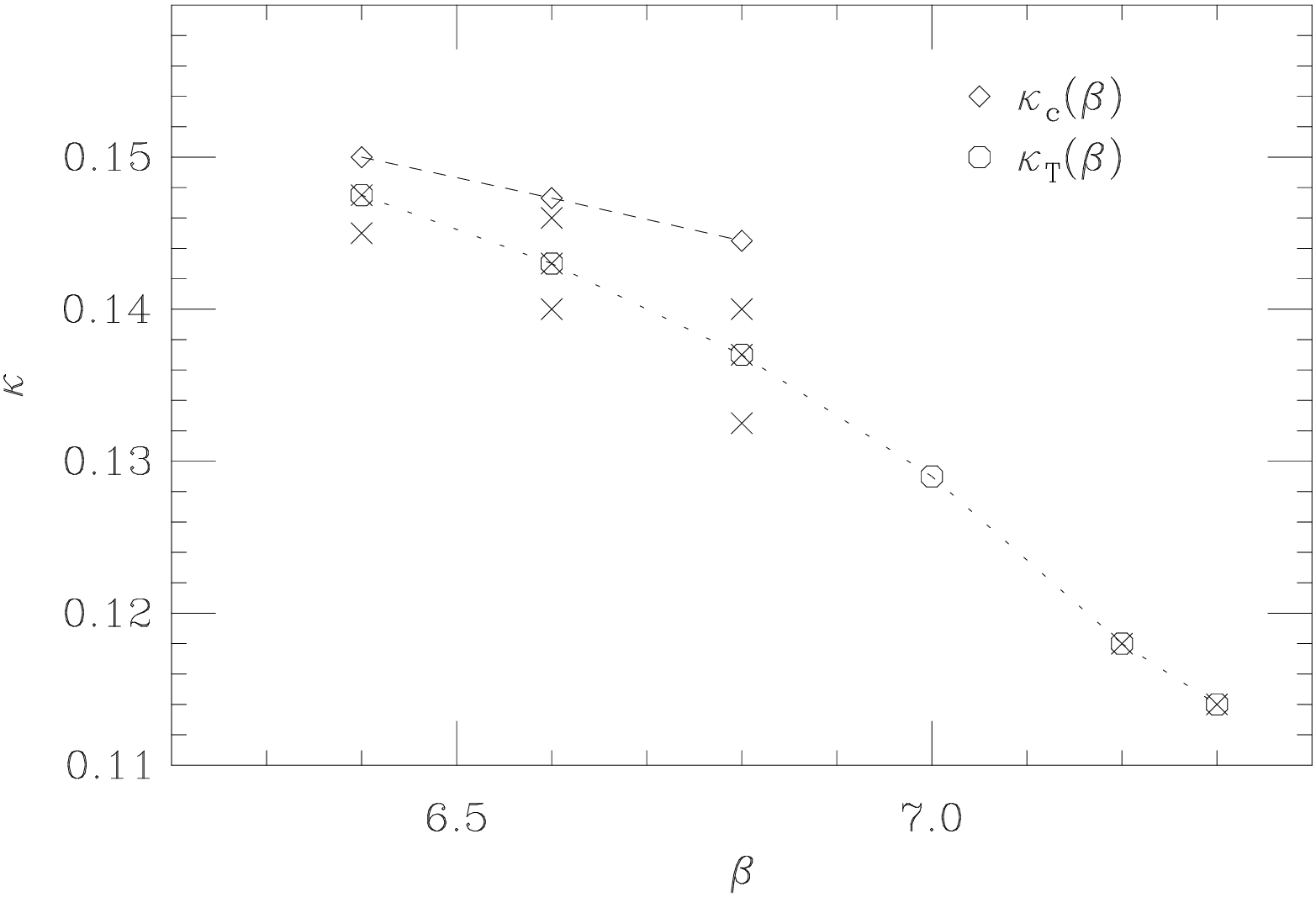}
\caption{ Phase diagram for the Symanzik-improved action.  Octagons represent
the $N_t=4$ thermal crossover, and diamonds indicate estimates of the
locations of vanishing pion mass.
Zero temperature simulations were performed at the crosses.
Dashed and dotted lines are merely to guide the eye. }
\label{fig:phase_cl}
\end{figure}

Our calculations of the hadron spectrum for our zero temperature
simulations are summarized in Table~\ref{tab:masses}, and
the phase diagram (Fig.~\ref{fig:phase_cl}) illustrates the
location of the $T=0$ runs with respect to the thermal crossover
and critical lines, $\kappa_T(\beta)$ and $\kappa_c(\beta)$
respectively.
An anomaly in Table~\ref{tab:masses} is the small data set for $\beta = 6.4$,
$\kappa = 0.1475$.  Naturally we would prefer to have more
configurations with which to compute hadron correlators.  Unfortunately
the cost of running at those parameters is high.

\begin{table}[hb]
\caption{\label{tab:masses} Masses of the pseudoscalar and vector 
mesons and the nucleon on an $8^3 \times 16$ lattice.
Points marked by as asterisk lie on the $N_t=4$ crossover.
The \# column lists the number of propagators used to compute
spectroscopy.}
\begin{center}
\begin{tabular}{rccc|ccccc}
$\beta$~ & $\kappa$ & $u_0$ & \# & $aM_{\rm PS}$ & $aM_{\rm V}$ & $aM_N$
& $M_{\rm PS}/M_{\rm V}$ & $M_N/M_{\rm V}$  \\ \hline
6.40 & 0.145 & 0.826 & 81& 0.931(4)& 1.351(18)& 2.14(3)& 0.689(10)& 1.58(3)\\
${}^*$6.40 & 0.1475 & 0.828 & 30& 0.664(8)& 1.26(6)& 1.63(11)& 0.527(26)& 1.29(11)\\
6.60 & 0.140 & 0.834& 64 & 1.173(4)& 1.481(9)& 2.30(3)& 0.792(6)& 1.55(2) \\
${}^*$6.60 & 0.143 & 0.841 &144 & 0.927(4)& 1.280(8)& 1.958(18)& 0.724(6)& 
1.530(15)\\
6.60 & 0.146 & 0.855& 40 & 0.468(15)& 1.04(13)& 1.34(5)& 0.45(6)& 1.29(17)\\
6.80 & 0.1325& 0.842& 79& 1.494(3)& 1.700(7)& 2.651(13)& 0.879(4)& 1.56(1)\\
${}^*$6.80 & 0.137 & 0.849& 120& 1.187(3)& 1.421(6)& 2.190(10)& 0.835(4)& 1.541(10)\\
6.80 & 0.140 & 0.857& 43& 0.885(8)& 1.182(16)& 1.75(4)& 0.749(12)& 1.48(4)\\
${}^*$7.20 & 0.118 &0.864 & 143 & 1.915(3)& 1.994(3)& 3.110(5)& 0.960(2)& 1.560(3)\\
${}^*$7.30 & 0.114 & 0.8695& 30& 2.043(4)& 2.106(5)& 3.297(11)& 0.970(3)& 1.614(6)
\end{tabular}
\end{center}
\end{table}

While in Figure~\ref{fig:pl_vs_k} we do not see the same first-order
jump in $\langle{\rm Re}~P\rangle$ that we did with the standard
Wilson actions, we would like to make the comparison more convincing.
After all, we cannot know {\it a priori} the relation between the
bare parameters for the standard action $(\beta_W,\kappa_W)$ and those
for the improved action $(\beta_I,\kappa_I)$; it could happen
that a small change in $\kappa_W$ corresponds to a much larger
change in the quark mass than does a similar change in $\kappa_I$,
giving us the illusion that the crossover is broader for the improved
action. 

\begin{figure}
\vspace{3.75in}
\includegraphics{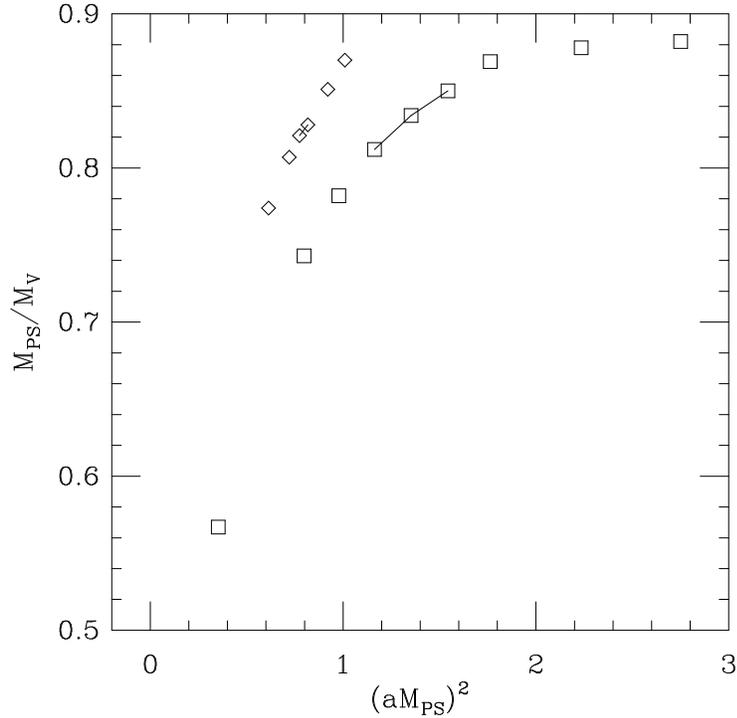}
\caption{Pseudoscalar/vector meson mass ratio vs.~the lattice
pseudoscalar mass squared.  The diamonds correspond to $\beta_W=4.9$
simulations with the unimproved Wilson action, and the squares 
denote our $\beta_{IW}=6.8$ simulations with the improved action.
The solid lines indicate the region of the thermal crossover. }
\label{fig:pirho_vs_mpisq2}
\end{figure}

\begin{figure}
\vspace{3.25in}
\includegraphics{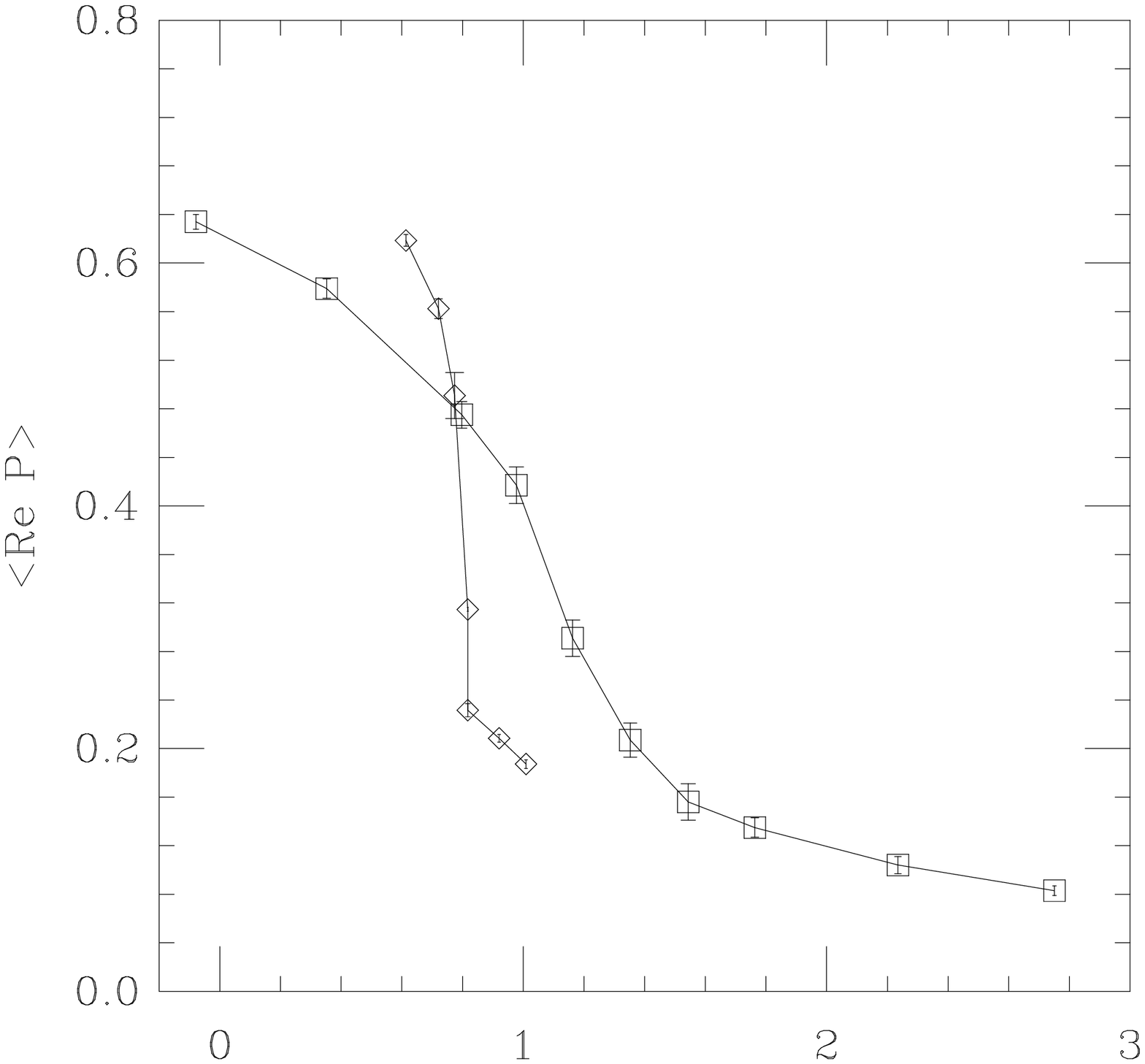}
\caption{Polyakov loop as a function of $(aM_{\rm PS})^2$ for fixed $\beta$.
Squares: $\beta=6.8$ improved Wilson fermions,
Diamonds: $\beta=4.9$ unimproved Wilson fermions.
Both have similar $M_{\rm PS}/M_{\rm V}$ at the crossover. }
\label{fig:rp_vs_mpisq2}
\end{figure}

Using measurements of the pseudoscalar mass near the crossover region, we
can interpolate in order to estimate $(aM_{\rm PS})^2$ as a function of
$1/\kappa$ for both actions.  In particular, we look at the thermal
crossover for improved and unimproved $N_t=4$ Wilson actions
at comparable $M_{\rm PS}/M_{\rm V}$.  Data from Ref.~\cite{ref:HTMCGC}
suggests that we compare $\beta_W=4.94$, $\kappa_W=0.18$ where
$M_{\rm PS}/M_{\rm V} = 0.836(5)$ to our $\beta_{IW}=6.80$,
$\kappa_{IW}=0.137$ where $M_{\rm PS}/M_{\rm V} = 0.835(4)$.
The data we use for comparison to ours are the unimproved $\beta_W=4.9$
data provided in Refs.~\cite{ref:MILC4,ref:SCRI_W}.
We interpolated the pseudoscalar meson mass using a linear least
squares fit to a quadratic in $1/\kappa$ around the crossover
region for the unimproved Wilson data.  Due to the smaller data sample,
we interpolated $(aM_{\rm PS})^2$ linearly in $1/\kappa$ for our
improved Wilson data.  Figure~\ref{fig:pirho_vs_mpisq2} is a plot
of $M_{\rm PS}/M_{\rm V}$ vs.\ $(aM_{\rm PS})^2$ for both actions.
The crossover occurs at similar $M_{\rm PS}/M_{\rm V}$ as indicated
by the solid lines in the graph.

Having eliminated bare quantities, we can plot the thermodynamic
observables against the pseudoscalar mass squared.  
Figure~\ref{fig:rp_vs_mpisq2} demonstrates that the crossover is indeed
smoother for improved action than the Wilson action.

\subsection{Heavy Quark Potential}
\label{ssec:hqp}

In addition to computing hadronic masses, we used Wilson
loop data to measure the heavy (or static) quark potential
$V(\vec r)$:
\beq
V(\vec r) ~=~ - \lim_{t\rightarrow\infty} ~{1\over t} ~\ln W(\vec r, t).
\eeq
A standard ansatz for the form of the potential is
\beq
V(r) ~=~ V_0 + \sigma r - {e\over r}
- f\bigg(G_L(r) - {1\over r}\bigg),
\eeq
where $V_0$, $\sigma$, $e$, and $f$ are fit parameters, and 
$G_L$ is the lattice Coulomb potential.  In practice, this fit
is performed for a fixed $t$; that is, the potential is estimated
through an effective potential,
\beq
V_t(\vec r) ~=~ -\ln \bigg[ {W(\vec r, t+1)\over W(\vec r, t)}\bigg],
\eeq
such that
\beq
W(r, t) ~\sim~ \exp(-V_t(r) t)
\eeq

The parameter $r_0$ is defined to be the length such that
\beq
r_0^2 F(r_0) ~=~ 1.65,
\eeq
with
\beq
F(r) ~=~ {\partial V \over \partial r},
\eeq
which corresponds to $r_0=0.49$ fm from potential models.
Sommer showed this to be a useful quantity with which
to set the lattice scale~\cite{ref:SOMMER}.  In this work
we calculated the force by taking numerical differences of the
potential.  Our analysis proceeds as in Ref.~\cite{ref:HELLER_HQP}.
Errors are estimated by bootstrapping the data, and occasionally
increased to account for differences in the choice of $t$.
We present our fits to the potential for the zero temperature
simulations along the $N_t=4$ crossover with improved Wilson
fermions in Table~\ref{tab:pot_sw}.

For comparison, we performed the same calculation with 
two flavors of Kogut-Susskind fermions for three parameter
sets along the $N_t=4$
crossover.  Our fits are given in Table~\ref{tab:pot_ks}.
Meson masses were taken from Table~1 of Ref.~\cite{ref:EOS4}.
In addition, we measured the potential at one point along the
$N_t=6$ KS crossover.  That fit also appears in Table~\ref{tab:pot_ks}.
The generation and spectroscopy of those configurations are discussed
in Ref.~\cite{ref:MILC_KS}.

\begin{table}
\caption{\label{tab:pot_sw} Fits to the heavy quark potential along the
$N_t=4$ crossover ($n_f=2$ improved Wilson fermions).}
\begin{center}
\begin{tabular}{ccc|cc|ccccc}
$\beta$ & $\kappa$& \# & $r_{\rm min}-r_{\rm max}$ & t &
$aV_0$ & $a^2\sigma$ & $e$ & $f$ & $r_0/a$ 
\\ \hline
6.40 & 0.1475& 30& 1.41-4.47& 2& 1.0(3)& 0.41(8)& 0.7(2)& 5.6(6)& 1.52(4)\\
6.60 & 0.1430& 108& 1.41-6.93& 2& 0.65(9)& 0.42(3)& 0.34(8)& 3.21(24)& 1.77(2)\\
6.80 & 0.1370& 95& 1.41-6.93& 2& 0.70(6)& 0.346(15)& 0.38(5)&2.45(18)&1.913(13)\\
7.20 & 0.1180& 117 & 1.41-5.66& 2&0.65(2)& 0.253(6)& 0.33(2)& 1.06(8)& 2.287(13)
\end{tabular}
\end{center}
\end{table}

\begin{table}
\caption{\label{tab:pot_ks} Fits to the heavy quark potential along the
$N_t=4({}^*6)$ crossovers ($n_f=2$ Kogut--Susskind fermions).}
\begin{center}
\begin{tabular}{ccc|cc|ccccc}
$\beta$ & $am_q$ & \# & $r_{\rm min}-r_{\rm max}$ & t &
$aV_0$ & $a^2\sigma$ & $e$ & $f$ & $r_0/a$ 
\\ \hline
5.2875 & 0.025& 55& 1.41-6.93& 2& 0.80(10)& 0.30(3)& 0.46(10)& 1.46(20)& 1.99(4)\\
5.3200 & 0.050& 67& 2.24-6.93& 2& 0.68(22)& 0.29(4)& 0.2(3)& 5.7(1.1)& 2.17(11)\\
5.3750 & 0.100& 90& 1.00-5.66& 3& 0.62(8)& 0.288(23)& 0.26(7)& 0.56(12)& 2.20(4)\\ 
\hline
${}^*5.415$ & 0.0125 & 280 & 2.24-6.71 & 3 & 0.76(2) & 0.130(5) & 0.36(2)&
1.0(2) & 3.14(5)
\end{tabular}
\end{center}
\end{table}

\subsection{Scaling tests}

In Sections~\ref{ssec:thermo} and \ref{ssec:spectrum}
we showed that thermodynamics with
the improved action
does not have the same artificial first order behavior that
unimproved Wilson thermodynamics does.  However, in order
to make physical predictions which can be compared with
results from Kogut-Susskind thermodynamics, we must make use
of the spectrum and potential computations described in the
preceding two sections.

\begin{figure}
\vspace{2.75in}
\includegraphics{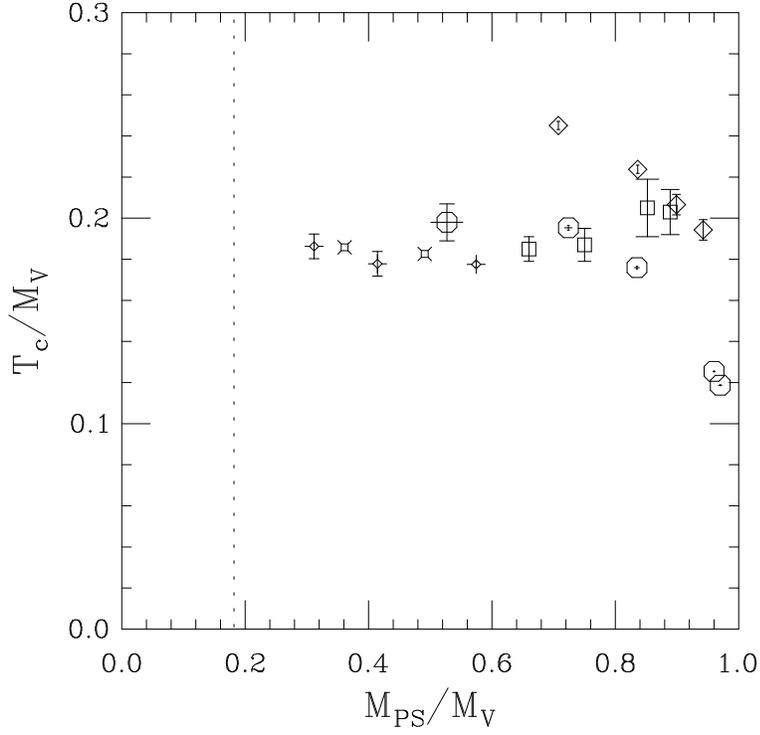}
\caption{Critical temperature divided by vector meson mass vs.
pseudoscalar/vector meson mass ratio.  Our data for $N_t=4$
improved Wilson actions are the octagons.  Diamomds: $N_t=4$
unimproved Wilson; Square: $N_t=6$ unimproved Wilson; Fancy
diamonds and squares: $N_t=4,6$ Kogut--Susskind (KS),
respectively~\protect\cite{ref:MILC6}.}
\label{fig:tc_over_mrho}
\end{figure}

In Figure~\ref{fig:tc_over_mrho} we plot the ratio 
$T_c/M_{\rm V}$ as a function of the pseudoscalar/vector
meson mass ratio $M_{\rm PS}/M_{\rm V}$.
Extrapolation to the physical pion/rho mass ratio is 
necessary in order to make a prediction for $T_c$.
The fact that $T_c/M_{\rm V}$ is independent of
$N_t$ for the Kogut--Susskind action leads one to believe
that this quantity is scaling at lattice spacing $a = 1/(4T_c)$.
Clearly, this statement is not true for the unimproved
Wilson action.  The $N_t=4$ unimproved Wilson points show a
large dependence on the quark mass, and disagree significantly
with the corresponding $N_t=6$ points at $M_{\rm PS}/M_{\rm V}
< 0.8$.  In addition, since $T_c/M_{\rm V}$ is
consistently lower for the improved action than the unimproved
action at equal lattice spacing, the discretization errors
in the latter must be appreciable.  The improved Wilson point
at $M_{\rm PS}/M_{\rm V}=0.53$ appears to have some slight
agreement with the Kogut--Susskind data, but with a large error.
Finally, we remark that
one expects $T_c/M_{\rm V}\rightarrow 0$ in the infinite
quark mass limit since the vector meson mass diverges there, so
ultimately we want to simulate at as small $M_{\rm PS}/M_{\rm V}$
as possible in order to extrapolate to $M_\pi/M_\rho=0.18$ reliably.

\begin{figure}
\vspace{3.25in}
\includegraphics{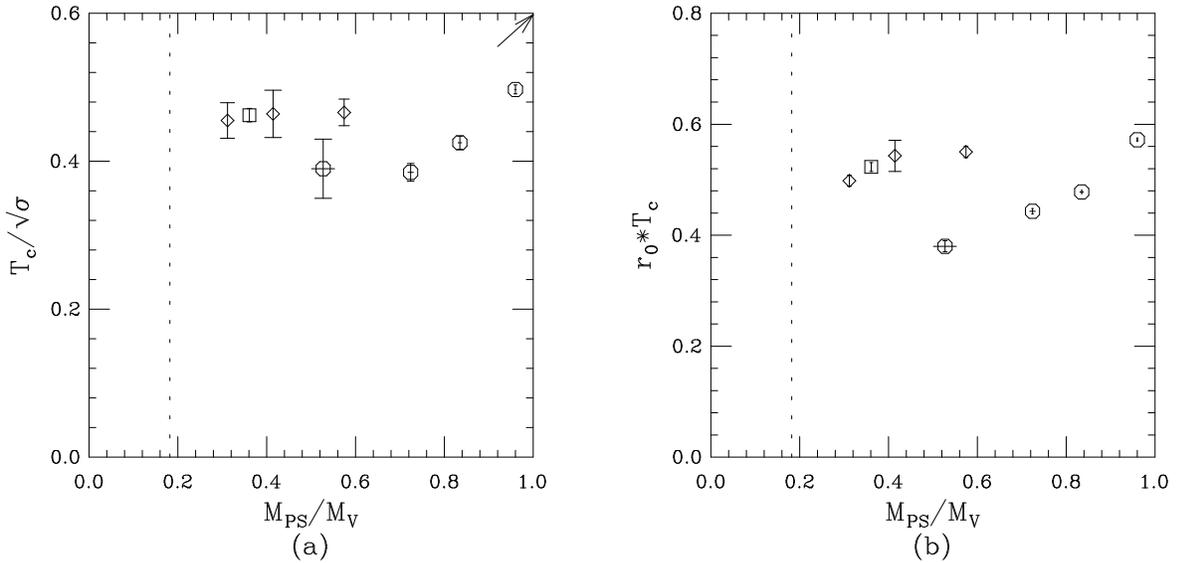}
\caption{Critical temperature scaled by (a) the square root of the
string tension and (b) the inverse Sommer parameter 
vs.~pseudoscalar/vector meson mass ratio.
Octagons: $N_t=4$ improved Wilson; Diamonds: $N_t=4$ KS; Square: $N_t=6$ KS.
The arrow in (a) shows the $N_t=4$ quenched
$T_c/\protect\sqrt{\sigma}$ 
from Ref.~\protect\cite{ref:BIELEFELD}.}
\label{fig:tc_pot}
\end{figure}

In order to look at $T_c$ scaled by quantities which are nominally 
independent of the quark mass, we use $\sqrt{\sigma}$ and $r_0$
from our potential fits mentioned in Sec.~\ref{ssec:hqp}.
In Figure~\ref{fig:tc_pot} the rise in $T_c/\sqrt{\sigma}$
and $r_0 T_c$ as $M_{\rm PS}/M_{\rm V}\rightarrow 1$ is presumably
due to $T_c$ approaching
the pure SU(3) transition temperature as the quarks decouple.
The $N_t=4$ quenched $T_c/\sqrt{\sigma}$ from Ref.~\cite{ref:BIELEFELD}
appears as an arrow in Fig.~\ref{fig:tc_pot} and supports this
presumption.  The disagreement between the Kogut--Susskind and
improved Wilson actions is more apparent in Fig.~\ref{fig:tc_pot}
than in Fig.~\ref{fig:tc_over_mrho}.  The error in $\sqrt{\sigma}$
is large, but both $T_c/\sqrt{\sigma}$ and $r_0T_c$ are lower for
our improved action than for the KS action.  In fact, the 
small error in $r_0$ reveals the presence of quark mass dependences
even at $M_{\rm PS}/M_{\rm V}=0.53$.

\begin{figure}
\vspace{2.25in}
\includegraphics{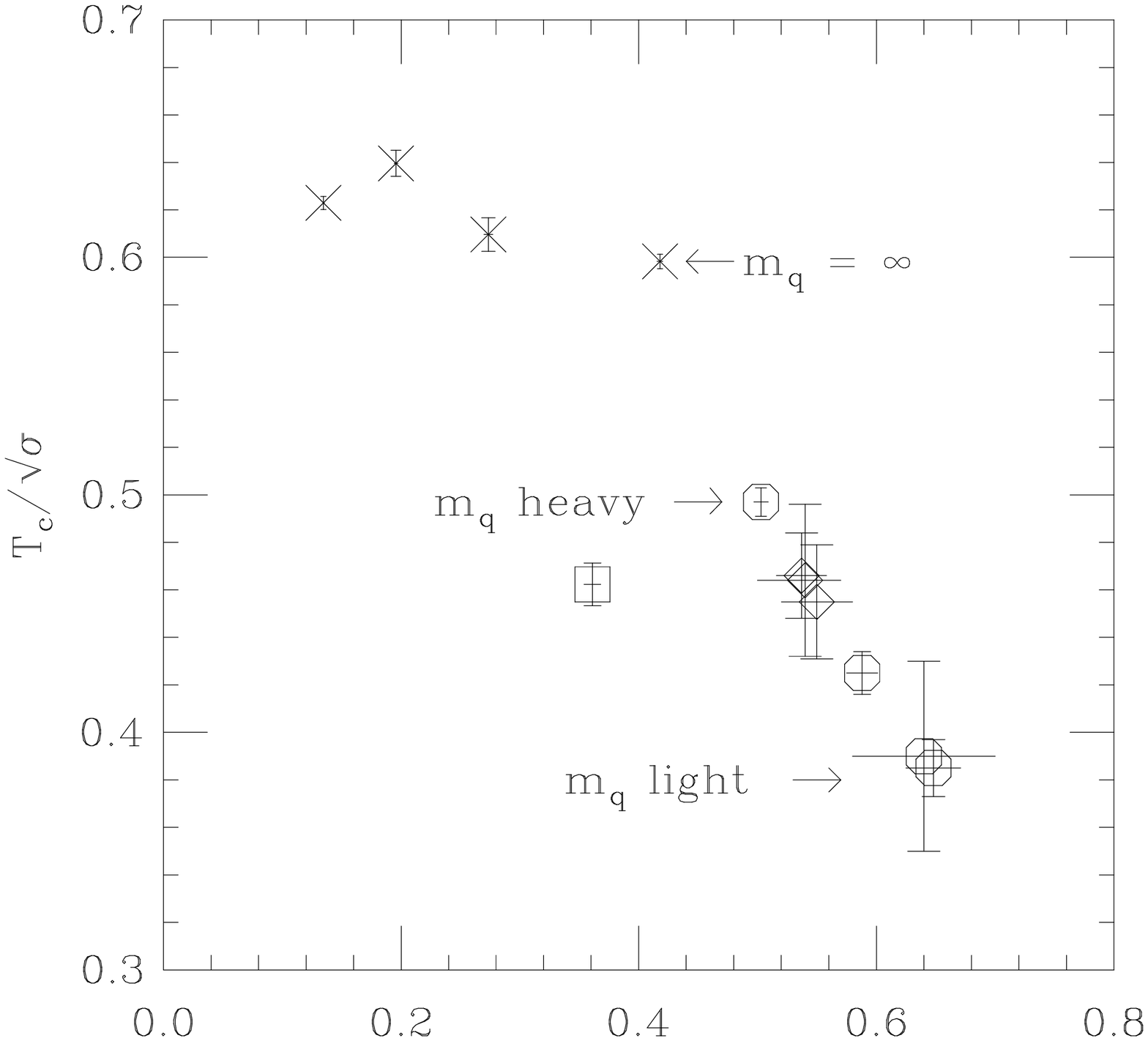}
\caption{Critical temperature scaled by the square root of the
string tension (left) vs.~the lattice spacing in units of 
$1/\protect\sqrt{\sigma}$.
Octagons: $N_t=4$ improved Wilson; Diamonds: $N_t=4$ KS; Square: $N_t=6$ KS;
Crosses: Quenched SU(3) from Ref.~\protect\cite{ref:BIELEFELD}.}
\label{fig:tcsigma_sigma}
\end{figure}

The quark mass effect can be identified 
further in the scaling plot of $T_c/\sqrt{\sigma}$ vs.\
$a\sqrt{\sigma}$ (Fig.~\ref{fig:tcsigma_sigma}).  Since $a = 1/(4T_c)$
for all of the $N_t=4$ data, the spread in $a\sqrt{\sigma}$ for
the improved action is caused by the increase in the deconfinement
temperature as the quarks become infinitely heavy.  One should contrast
to this the observation that the 3 $N_t=4$ KS points lie on top of each other.
The higher $N_t$ points for both KS and quenched actions show their
relative independence on lattice spacing.
The conclusion one should draw from Figs.~\ref{fig:tc_pot}
and~\ref{fig:tcsigma_sigma} is that in the case of the improved Wilson
data, any attempt at extrapolation to physical quark mass is premature.

\begin{figure}
\vspace{3.75in}
\includegraphics{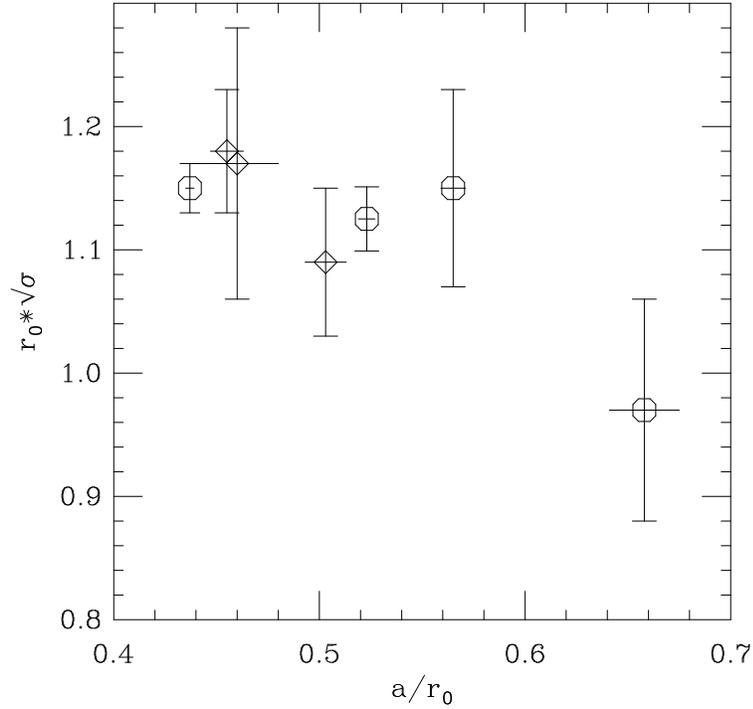}
\caption{The dimensionless quantity
$r_0\protect\sqrt{\protect\sigma}$ vs.~the
lattice spacing in units of $r_0$.
Octagons: $N_t=4$ improved Wilson, Diamonds: $N_t=4$ KS. }
\label{fig:r0sigma}
\end{figure}

In Figure~\ref{fig:r0sigma} we plot $r_0\sqrt{\sigma}$ vs.\
$a/r_0$.  While $r_0$ and $\sqrt{\sigma}$ scale together within
the error bars, the variation in $a/r_0$ for the
improved action along the crossover is another manifestation of
the scaling violations between the critical temperature and the
gluonic observables.  A plot against $a\sqrt{\sigma}$ looks
qualitatively the same, but with larger errors.

\begin{figure}
\vspace{3.25in}
\includegraphics{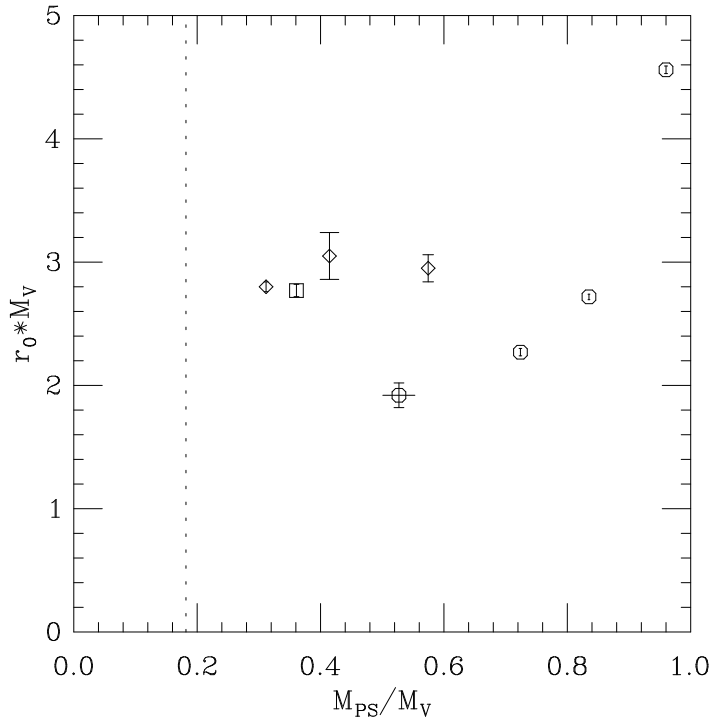}
\caption{Vector meson mass times $r_0$ vs.~pseudoscalar/vector
meson mass ratio.
Octagons: $N_t=4$ improved Wilson, Diamonds: $N_t=4$ KS, Square: $N_t=6$ KS. }
\label{fig:mrho_r0}
\end{figure}

A graph of the vector meson mass times $r_0$ (Fig.~\ref{fig:mrho_r0})
shows nice behavior for the Kogut--Susskind simulations, disagreement
between KS and clover, and the rise in $M_{\rm V}$ toward 
infinity at large $M_{\rm PS}/M_{\rm V}$.  Again, we do not show
$M_{\rm V}/\sqrt{\sigma}$ vs.\ $M_{\rm PS}/M_{\rm V}$ since it
is qualitatively the same, but with larger error bars.
If we were so bold as to argue that the improved Wilson data 
could be extrapolated to physical $M_\pi/M_\rho$ using the 
points $M_{\rm PS}/M_{\rm V} \leq 0.8$, then we
would conclude $M_{\rm V}/\sqrt{\sigma}$ for our action is less
than for the Kogut--Susskind action.
This would not be too surprising given similar trends in scaling
violations for quenched QCD spectroscopy as presented in Figure~2 of
Ref.~\cite{ref:SCRI_CLOV}, for example.  At finite lattice spacing, 
$M_\rho/\sqrt{\sigma}$ computed with KS valence quarks lie above
the $a = 0$ extrapolation, while unimproved Wilson quark calculations
give a value less than the continuum number.  The addition of the
clover term significantly reduces this scaling violation; however,
the lattice value of $M_\rho/\sqrt{\sigma}$ still lies below its
continuum value.  Of course, in the absence of clear scaling
between $M_{\rm V}$ and $\sqrt{\sigma}$ (and $r_0$), such arguments
in this work are speculative.

\section{Conclusions}
\label{sec:concl}

This is the first large scale simulation of unquenched QCD
with improved Wilson fermions of which we are aware.
We find that the Symanzik improvement
program, at this level, fulfills its promise in that a serious
lattice artifact, the spurious first-order transition at intermediate
hopping parameters, has been removed.  The thermal crossover does become 
progressively steeper as one decreases the quark mass, but it is
smooth in the sense that the Polyakov loop and the plaquette are
single-valued for all $(\beta,\kappa)$ at which we computed.

However, improvement at this order is no panacea.  It is still very
costly to invert the fermion matrix near and below
$M_{\rm PS}/M_{\rm V} \approx 0.5$.  Since the critical temperature
and the vector meson mass show a significant dependence on the quark
mass, extrapolation to $M_\pi/M_\rho$ is not trustworthy.
Furthermore, disagreement is evident in $T_c/\sqrt{\sigma}$ and
$M_{\rm V}r_0$ between our improved Wilson action
and the unimproved Kogut--Susskind formulation, even at
comparable $M_{\rm PS}/M_{\rm V}$.

One cannot yet use this disagreement to cast doubt on the 
Kogut--Susskind results because the scaling behavior of the
improved Wilson action has not been sufficiently tested.  
Simulations with smaller lattice spacing, perhaps $N_t=6$, 
would give a more concrete picture of the extent of scaling 
violations in this action.

Before beginning such an expensive undertaking, however, let us
speculate as to the shortcomings of the present action in the
context of thermodynamics.  In the high temperature phase,
thermodynamic quantities are dominated by high momentum contributions.
Therefore, one must not only improve the effects of the finite
lattice spacing, but also the dispersion relation at all momenta.
Although the gauge action we used has a dispersion relation closer
to the continuum than the plaquette action, the clover term does
not change the fermionic dispersion relation from that of the unimproved
Wilson action.  The work with an improved gauge action but standard
Wilson fermions by Ref.~\cite{ref:TSUKUBA} shows improvement similar
to ours: {\it viz.}\ removal of the jump-discontinuity in the
Polyakov loop.  A detailed comparison of the critical temperature
from their action versus ours and the standard Wilson and KS actions
remains to be made.  

Therefore, it is plausible that improvement of the gauge action is
responsible for the removal of the artificial first-order behavior
at intermediate values of the Wilson hopping parameter.  However,
improvement of the fermion action probably plays a role in the
closer agreement to the KS results for $T_c/M_{\rm V}$, as was
found in studies of quenched spectroscopy.  Persistent
quark mass dependence and apparent disagreement between our results
and KS results for $T_c$ scaled by quark potential parameters 
indicate that further improvement in the fermionic sector is warranted.
One might consider using Wilson-type fermions with
an improved dispersion relation in the next large scale thermodynamics
study.

\section*{Acknowledgments}

This work was supported by the U.S. Department of Energy under contracts
DE-AC02-76CH-0016,
DE-FG03-95ER-40894,     
DE-FG03-95ER-40906,
DE-FG05-85ER250000,
DE-FG05-96ER40979,      
DE-2FG02-91ER-40628,
DE-FG02-91ER-40661,
and National Science Foundation grants
NSF-PHY93-09458,
NSF-PHY96-01227,
NSF-PHY91-16964.
Simulations were carried out at the Cornell Theory Center,
the Supercomputer Computations Research Institute
at Florida State University, and at the San Diego Supercomputer Center.

One of us (MW) would like to extend thanks to A.~Hasenfratz for several 
helpful discussions and to N.~Christ, F.~Karsch, and A.~Ukawa for 
thoughtful comments at the Lattice '96 symposium.
We also thank Craig McNeile and Tom Blum for critical readings of the
manuscript.


\end{document}